\documentstyle[prl,aps,epsfig,amssymb]{revtex}
\def\B.#1{{\bbox{#1}}}
\def\x{{\mbox{\boldmath$x$}}}

\def\r{{\mbox{\boldmath$r$}}}

\def\unitr{{\mbox{\boldmath$\hat r$}}}

\def\unitz{{\mbox{\boldmath$\hat z$}}}

\def\v{{\mbox{\boldmath$v$}}}
\def\unitr{{\mbox{\boldmath$\hat r$}}}
\def\unitz{{\mbox{\boldmath$\hat z$}}}

\def\la{{\langle}}
\def\ra{{\rangle}}

\def\nab{{\bf \nabla}}

\newcommand{\rb}{Rayleigh-B\'enard}
\newcommand{\lp}{\left(}
\newcommand{\rp}{\right)}
\newcommand{\be}{\begin{equation}}
\newcommand{\ee}{\end{equation}}
\newcommand{\bea}{\begin{eqnarray}}
\newcommand{\eea}{\end{eqnarray}}

\begin{document}
\twocolumn[\hsize\textwidth\columnwidth\hsize\csname
@twocolumnfalse\endcsname
\title{Evidences of Bolgiano scaling in 3D \rb\ convection}
\author{Enrico Calzavarini$^{1}$, Federico Toschi$^{2}$ and Raffaele Tripiccione$^{1}$}
\address{$^1$ Universit\`a degli Studi di Ferrara, and INFN, Sezione di Ferrara,
I-34100 Ferrara, Italy.}
\address{$^2$ IAC-CNR, Viale del Policlinico 137, I-00161, Roma, Italy and\\
INFM, Unit\'a di Tor Vergata, Roma, Italy.}

\maketitle
\begin{abstract}

We present new results from high-resolution  high-statistics direct numerical
simulations of a tri-dimensional convective cell. We test the fundamental 
physical picture of the presence of both a Bolgiano-like and a Kolmogorov-like 
regime. We find that the dimensional predictions for these two
distinct regimes  (characterized respectively by an active and passive 
role of the temperature field) are consistent with our measurements.

\vskip0.2cm
\end{abstract}

%PACS: 47.27-i, 47.27.Nz, 47.27.Ak
\vskip0.3cm
\vskip0.2cm
]

%%%%%%%%%%%%%%%%%%%%%%% 

\noindent The dimensional theory of homogeneous and isotropic turbulence has
been definitely settled long ago by the work of A. N. Kolmogorov
\cite{frisch}.  On the other hand, it is still missing a clear theoretical picture  
for the strong fluctuations in the energy dissipation field that lead to
intermittency effects (i.e. non gaussian behaviour of probability distribution
functions). Phenomenological theories have been proposed \cite{frisch} but no
systematic theory for computing experimentally measured numbers has been
successful so far. The situation of ``non-ideal'' turbulence is even more
controversial, already at the level of dimensional expectations. A typical
realization, the one we will address in this paper, is the tri-dimensional (3D)
\rb\ cell, described, in the Boussinesq approximation \cite{landau}, by the
following set of equations:
\bea
\label{eq1} \partial_t \v &+& \lp\v\cdot\nab\rp \v = -\nab p + \nu \nabla^2 \v + \alpha g T \unitz\\
\label{eq2} \partial_t T &+& \lp\v\cdot\nabla\rp T =  \chi \nabla^2 T
\eea
\noindent
with isothermal boundary conditions on the upper and lower planes of a cell of
height H: $T(z=0)=+\Delta T/2$ and $T(z=H)=-\Delta T/2$.  As usual, $\v(\x,t)$
is the velocity field and  $T(\x,t)$ the temperature field. Kinematic viscosity
and  thermal conductivity are respectively  $\nu$ and $\chi$, while the thermal
expansion coefficient is $\alpha$ and gravity acceleration is $g$. In the
following we will mainly focus on the longitudinal structure functions of $v$
and $T$:  $S_p(r)=\la\left[\lp \v(\x+\r)-\v(\x) \rp \cdot \unitr \right]^p \ra$
and  $T_p(r)=\la \left[ T(\x+\r)-T(\x) \right]^p \ra$.\\

In this work we present some tests of the predictions for the structure
functions defined above
that can be derived in the scenario proposed years ago by
Bolgiano \cite{bolgiano} to describe convective turbulence.

Despite much research on the subject, sound evidence of the validity of the Bolgiano
scenario and the recovery of Kolmogorov scaling at small scales is still missing. Good
quality  confirmation of the dimensional Bolgiano scenario was recently shown in a two
dimensional numerical simulations \cite{2D}. This result cannot be directly related to
the 3D case because of the strong differences in the properties of the velocity field in 2D.
Statistical properties of the velocity field were recently used with even simpler
models (shell models for turbulence)  \cite{goy}. This approach is even less probing
since these models  were built  precisely in order to implement Bolgiano scaling.
Finally, a recent experiment even questioned the  behaviour the the Bolgiano
length inside a convective cell \cite{zhou01}.  Of course from an experimental point
of view it can be difficult, if not impossible at all, to measure all relevant
quantities and it can be even harder to have access to information at several (not
just a few) positions inside the volume.
The results presented in this letter show good consistency with the idea of a Bolgiano
regime (i.e. a range of length scales where temperature driven buoyancy effects are dominant) but, because of the still limited resolution of ``state of the art''
numerical simulations,  we will have to resort to somehow indirect tests.

This letter is organized as follows: a brief review of phenomenological expectations, 
details of our numerical simulations, data analysis and then concluding remarks.\\

Starting from equations (\ref{eq1}) and (\ref{eq2}), if one uses dimensional analysis
and assumes homogeneous scaling for velocity and temperature differences (inside the inertial range),
one ends up with two distinct scaling regimes. 
At small scales (Kolmogorov-like scenario),
$r\ll L_B$,

\bea
\label{sotto1} \delta v(r) &\sim& \varepsilon^{1/3} r^{1/3}\\
\label{sotto2} \delta T(r) &\sim& N^{1/2} \varepsilon^{-1/6} r^{1/3}
\eea
while at large scales (Bolgiano-like scenario), $r\gg L_B$,
\bea
\label{sopra1} \delta v(r) &\sim& \lp\alpha g\rp^{2/5} N^{1/5} r^{3/5}\\
\label{sopra2} \delta T(r) &\sim& \lp\alpha g\rp^{-1/5} N^{2/5} r^{1/5}
\eea

\noindent
The Bolgiano length, $L_B$, is an estimate of the distance at which 
the dissipative and buoyancy terms on the right hand side of equation 
(\ref{eq1}) balance, valid under the assumption of a scaling behaviour.
In the following we will use the $z$ dependent version of
$L_B(z)$ introduced in \cite{jsp} that uses averages of $\varepsilon$
and $N$ defined at a given height $z$,
$\varepsilon (z) = {\nu /2}\la \sum_{ij} \lp{\partial_i v_j}\rp^2\ra_z$ 
and  $N(z)={\chi /2} \la \sum_i \lp\partial_i T\rp^2\ra_z$:

\be
\label{lb}
L_B(z) = {{\varepsilon(z)^{5/4}} \over {N(z)^{3/4} {(\alpha g)}^{3/2}}}
\ee

\begin{figure}[!ht]
\hskip -.4cm\epsfig{file=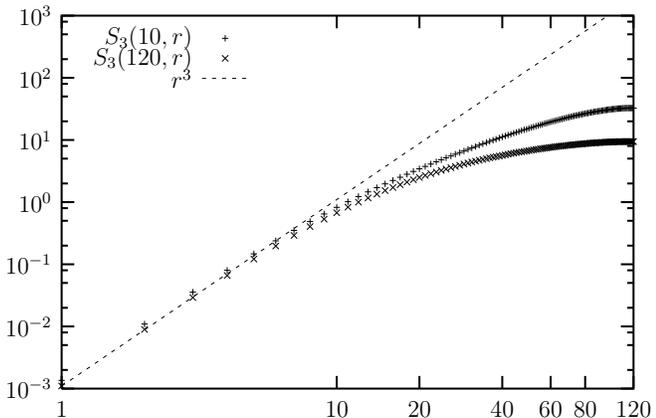,width=\hsize}
\caption{\label{fig0}Log-log plot of the structure function $S_3(r)$, 
defined in the text, measured close to the end plates ($+$) and at the center of the cell ($\times$). The horizontal scale is in grid points, while the vertical scale is in arbitrary units.}
\end{figure}

Our analysis has been performed on data coming from Direct Numerical
Simulations (DNS) employing a by now standard Lattice  Boltzmann scheme
\cite{lb}, on a massively parallel computer\cite{ape}.
The resolution of the numerical simulation was $240^3$ and the
Rayleigh number ($Ra = {{\alpha g \Delta T H^3} / {\lp \nu \chi\rp}}$) was approximately  
$ 3.5 \cdot 10^7$. The Prandtl number was  equal to unity. We
performed a stationary simulation extending over approximately
$500$ recirculation times
and stored nearly $400$ independent configurations
with full information on all velocity components and the temperature field.
Boundary conditions were periodic in the $x$ and $y$ directions (in order to
maintain homogeneity on horizontal planes) and isothermal at the top and bottom
planes of the cell ($z=0$ and $z=H$). 

Free slip boundary conditions (i.e. satisfying only the incompenetrability
constraint) were used on the top/bottom planes for the velocity field.  This
choice was made in an attempt to reduce the effects of 
a viscous  boundary layer close to the horizontal walls.  Indeed, as
will be clear from the following, thermal effects are dominant near the
isothermal walls, so using no-slip boundary conditions might produce effects on
the velocity statistics interfering with the ones coming from pure buoyancy.
More details on this point will be given later on.

The most direct way to test the dimensional validity of the Bolgiano picture
would be to measure structure functions and to  check whether they scale
with exponents close (apart from intermittency corrections)
to the ones predicted  by the sets of equations (\ref{sotto1}-\ref{sotto2}) and
(\ref{sopra1}-\ref{sopra2}).

Unfortunately this cannot be done directly because of the limited resolution 
of our DNS.

Fig. \ref{fig0} substantiates this comment, by plotting the
velocity structure function of order 3, $S_3(z,r)=\la \lp v(x+r,z)-v(x,z)\rp^3
\ra$ at two different position: one very close to the wall ($z=10$), and
the other
at the center of the cell ($z=120=H/2$). 

\begin{figure}
\hskip -.4cm\epsfig{file=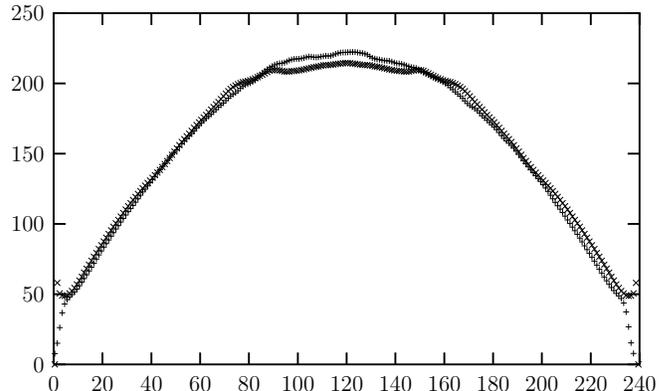,width=\hsize}
\caption{\label{fig1}Plot of  the z-dependent Bolgiano length as a function of
the distance  from the bottom wall, $z$, as given by the usual definition of
eq. \ref{lb} ($L_B(z)$, $+$) and as given by the procedure described in the
text,
following eq. \ref{ourlb} ($\alpha \cdot {\tilde L}_B(z)$, $\times$). 
All dimensions are in grid points and $\alpha = 3.1$.}
\end{figure}

As it can be seen from Fig.
\ref{fig0}, no evident scaling range can be detected, even if a steepening of
$S_3(r)$ is clearly detected as one comes close to the walls.\\

Given this state of affair, in order to test the consistency with the Bolgiano picture we
performed two  distinct, even if somehow less direct, tests.\\
The first test consists in checking that the Bolgiano length actually
keeps track of the scales at which  the buoyancy term balances the dissipative term
in equation (\ref{eq1}).

First, we measure $L_B(z)$ in terms of $\varepsilon(z)$ and $N(z)$ and plot our
results in Figure \ref{fig1}. From a first look at the  behaviour of $L_B(z)$
we learn that Bolgiano effects should be measurable, if at all, near to the
isothermal walls (where $L_B(z)$  is of the order of $10^1$).

Close to the center of the cell, $L_B$ is of the order of $10^2$ and Kolmogorov
like behaviour is expected to be measurable at almost all scales.

We then measure, directly and independently from the previous quantity, the
scale at which dissipation and buoyancy effects balance, i.e. we look for the
scale ${\tilde L}_B(z)$ for which:

\be
\label{ourlb}
\varepsilon(z) \sim \alpha g \la \delta v({\tilde L}_B)\delta T({\tilde L}_B) \ra_z
\ee

We provisionally consider ${\tilde L}_B$ a modified definition of the Bolgiano
length. In Fig. \ref{fig1} we plot  $L_B(z)$ and $\alpha {\tilde L}_B(z)$ with a 
constant which was tuned to be $\alpha\sim 3.1$.

As it can be seen, the two definitions yield the same behaviour apart
from the multiplicative factor, $\alpha$.   The reason for the multiplicative factor 
(of order unity)  is due to the fact that the two definitions 
are dimensional estimates so they can miss a numerical prefactor. Considering this, 
the fact that the two definition behave in the same way after rescaling has to be considered, 
in our opinion, an excellent agreement.

Here we want to underline two points which, we believe, add relevance to this finding.
First of all the two definitions are of course linked but definitely different.
The ``traditional'' definition of $L_B$ (see also \cite{jsp}), as from definition (\ref{lb}), 
comes from supposing that the two scaling laws in equations (\ref{sotto1}) and (\ref{sopra1}) 
merge at $L_B$, hence comes from solving the equation
$\varepsilon(z)^{1/3} L_B(z)^{1/3} = \lp \alpha g \rp^{2/5} N(z)^{1/5} L_B(z)^{3/5}$.
The second definition, ${\tilde L}_B(z)$ comes instead from a {\it direct} measurement of the
strength of the dissipative and forcing term in equation (\ref{eq1}).

The second important point consist on the fact that our cell is {\em not}
homogeneous; this  adds strength to the equality between the two different
definitions of $L_B(z)$.\\

As a consequence of this test we can claim that  the Bolgiano
scenario and the expected power law behaviour are consistent with the value
measured for the terms appearing on the right hand side of equation
(\ref{eq1}).  Furthermore we fully confirm (with higher accuracy) our former
results for $L_B$ \cite{jsp}.  We like to underline that extracting the
behaviour of the Bolgiano length using measured quantities that are not those
appearing in its definition could introduce large errors.
For example in \cite{zhou01} a Bolgiano length was
extrapolated as the scale at which there is a  change of slope in a particular
structure function.  
In using such a procedure one is dominated by strong finite size effects
present on the  structure functions.\\

We now proceed to our second test. Under the
hypothesis of validity of Bolgiano scaling, and if enough resolution were
available, one would expect to see power law behaviour in the inertial range
in Fig. \ref{fig0} with two distinct slopes $1$ and $9/5$, corresponding
respectively to Kolmogorov and Bolgiano scaling.
With available resolution, we are not able to detect a clear power law
behaviour from Fig. \ref{fig0}, although we see a clear
steepening of the structure function as we come close to the wall.
What we are going to do in the following is to try to quantify as well as
possible this change of slope.

We adopt the following procedure. We focus on the plot of $S_6(r)$ versus
$S_3(r)$ and apply Extended Self Similarity (ESS, see \cite{ess}) in order to detect a trustable
plateau in the local slopes. We found this plateau to correspond roughly to
distances in the interval ${\cal I}^{v} = [25,40]$ for the velocity structure function
 and ${\cal I}^{T}= [15,30]$ for the structure functions of the
temperature field.
We then define two other interval slightly shifted to the left ${\cal
I}^{v}_{-} = [20,35]$, ${\cal I}^{T}_{-} = [10,25]$ and to the right ${\cal
I}^{v}_{+} = [30,45]$, ${\cal I}^{T}_{+} = [20,40]$.
These interval were of course shifted by a reasonable amount, i.e. 
were still possible highest or lowest estimate for the same plateau. 

We finally perform a power law fit to extract a scaling exponent on
the structure functions for the velocity and for the temperature in the  three
intervals defined before.
\begin{figure}[!b]
\hskip -.4cm\epsfig{file=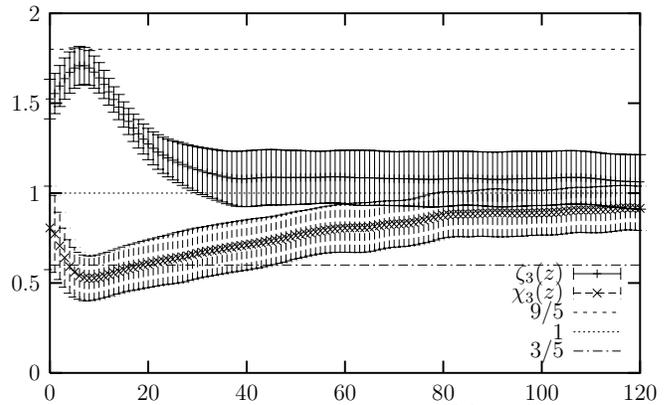,width=\hsize}
\caption{\label{fig2}Behaviour of the exponents $\zeta_3(z)$, $\chi_3(z)$ as a function of $z$.}
\end{figure}

We define the scaling exponents for the structure functions of interest as follows:

\bea
\label{exp1}\la \delta v(z,r)^p\ra &\sim& r^{\zeta_p(z)}\\
\label{exp2}\la \delta T(z,r)^p\ra &\sim& r^{\chi_p}(z)\\
\label{exp3}\la \lp \delta v(z,r) \delta T(z,r)^2 \rp ^{p\over 3}\ra &\sim& r^{\rho_p(z)}
\eea
Our fits provide a central value for the exponents and two, respectively,
higher and lower estimates (corresponding to the shifted intervals). 
The procedure adopted in estimating the errors
reflects the fact that the largest source of systematic error is connected
with the choice of the fitting ranges and not with statistical accuracy.

In the Kolmogorov regime we expect the exponents of equations
(\ref{exp1}-\ref{exp3}) to take the following dimensional values: $\zeta_p =
p/3$, $\chi_p = p/3$ and $\rho_p=p/3$. In the Bolgiano regime, on the other
hand,  we expect the following dimensional values: $\zeta_p = 3/5p$, $\chi_p =
p/5$ and $\rho_p=p/3$.

In Figure \ref{fig2} we plot the behaviour of the measured central exponents
and their errors as a function of $z$ for $\zeta_3(z)$ and $\chi_3(z)$. The 
behaviour  is qualitatively and quantitatively consistent with  the expected scaling exponents: 
we observe a smooth transition from a Bolgiano dominated regime (near to the wall, 
small $z$) to a Kolmogorov regime (approaching the center of the cell).

In Figure \ref{fig2} we have not shown the behaviour of $\rho_3(z)$ as it 
is consistent (within error bars) with the constant value $1$ and plotting
it on the same figure would have made it unreadable.

Here we, once again, like to underline the consistency of our findings. From the theoretical
picture we expect to see the crossover to the Kolmogorov scaling when
$L_B(z)\le H$. This is indeed the case as $L_B(z)\le H$ for $z\le 40$
and indeed we see $\zeta_3(z)\sim 1$ in the same range of $z$.

Another interesting question concerns the ``real'' statistics of the Bolgiano
regime. 
Very recently interest is growing along this line of research because of the desire to understand  the differences of the statistical properties of an active with respect to a passive scalar \cite{2D,goy}.
In this work we  focused on the gross features, i.e. on the
dimensional behaviour. 
If we try to look at intermittency, by means of ESS we find a strong increase of intermittency for the velocity field approaching the isothermal walls.

Unfortunately we believe that with our simulation we are not in a position to make any definite statements about intermittency in the Bolgiano dominated regime.

Indeed a study of intermittency in the Bolgiano regime would involves positions
nearby the isothermal walls (as only there the Bolgiano length scale is small
enough to have an inertial range largely dominated by buoyancy effects). 
Recently it was found that intermittency increases in the velocity structure
functions inside a viscous boundary layer \cite{myprl}.  In an attempt to
reduce the viscous boundary layer  we decided to apply free-slip velocity
boundary conditions to the isothermal walls free slip for the velocity.
Unfortunately our idea cannot be successful because of two reasons. First, the
free-slip boundary conditions does not completely suppress the boundary layer.
A boundary layer thickness can be defined through an extrapolation of $\la
v_z(x)^2\ra$, and in our case turned out to be of the order of roughly $15$
grid spacings. Second, it was recently realized that mechanical turbulence in a
free-slip channel also presents an enhancement of intermittency near to the
walls \cite{schumacher}.

A procedure to disentangle buoyancy and planar effect is clearly needed to make
any definite statement on intermittency in the Bolgiano regime. Otherwise one cannot
decide whether a change of intermittency is related to Bolgiano dynamics or to the increase
occurring nearby boundary layers.

In order to clarify this point it would be important to perform a simulation of
a \rb-like system with periodic boundary condition in all directions (i.e. an
homogeneous \rb\ cell).  We suggest to extend to 3D the study made in 2
dimensions in
\cite{2D}.

In order to do that one could write temperature field as the sum of a linear
 profile plus a fluctuating part, 
$T(x,y,z)=T_{\mbox{\tiny lin}}(z) + T'(x,y,z)$, with 
$T_{\mbox{\tiny lin}} = \Delta T/2 \cdot (1-2z/H)$ and obtain:

\bea
\label{omeq1} \partial_t \v &+& \lp\v\cdot\nabla\rp \v = -\nabla p + \nu \nabla^2 \v + \alpha g T {\unitz}\\
\label{omeq2} \partial_t T' &+& \lp\v\cdot\nabla\rp T' =  \chi \nabla^2 T' - {{\Delta T} \over H} v_z
\eea

Choosing the parameters in order to have a Bolgiano length as small as
possible, one would  benefit of a wide range of scales were to study the
buoyancy dominated flow. Further advantage of homogeneity would be the natural
increase of statistics and also  the applicability of tools like $SO(3)$
decomposition, to disentangle anisotropic terms \cite{so3_1,so3_2}.  
A study of this kind is in progress.

Concluding we have performed a number of basic tests in order to validate
the scenario of Bolgiano-Kolmogorov scaling in a convective cell, within the
limitations but also advantage of $z$ non homogeneity of our cell. We were
able to confirm the transition between the two expected scenarios.

We acknowledge useful discussions with R. Benzi, S. Succi and R. Verzicco. All
numerical simulations were performed on the APEmille computer at INFN, Sezione
di Pisa.

% in part by the EU under the Grant No. HPRN-CT  2000-00162 ``Non Ideal Turbulence'' and by INFN, Pisa.
%%%%%%%%%%%%%%%%%%%%%%%%

\end{document}